\documentclass[twocolumn]{article}
\usepackage{graphicx} 
\usepackage{cite}
\usepackage{amsmath,amssymb,amsfonts}
\usepackage{graphicx}
\usepackage{textcomp}
\usepackage{chapterbib}  
\usepackage{caption}
\usepackage{subcaption}
\usepackage{array}
\usepackage{multirow}
\usepackage{caption}
\usepackage{subcaption}
\usepackage{booktabs}
\usepackage{multirow}
\usepackage{hyperref}
\usepackage[inkscapelatex=false]{svg}
\usepackage{pdfpages}
\usepackage{geometry}
\usepackage{algorithm}
 \usepackage{float}
 \usepackage{comment}
\usepackage{placeins} 
 \usepackage{amsmath}
\usepackage{algorithm}
\usepackage{algpseudocode}

\usepackage{graphicx}
\usepackage{array}
\usepackage{authblk}
\makeatletter
\newcommand\footnoteref[1]{\protected@xdef\@thefnmark{\ref{#1}}\@footnotemark}
\makeatother
\title{Generative augmentations for improved cardiac ultrasound segmentation using  diffusion models}
\author{Gilles Van De Vyver, Aksel Try Lenz, Erik Smistad, Sindre Hellum Olaisen, Bjørnar Grenne, Espen Holte, Håvard Dalen, and Lasse Løvstakken}
\date{}
\begin{document}

\maketitle

\begin{figure*}[h!]
\centering
  \centering
  \includegraphics[trim={0.75cm 0.25cm 0.75cm 0.25cm}, clip,width = \linewidth]{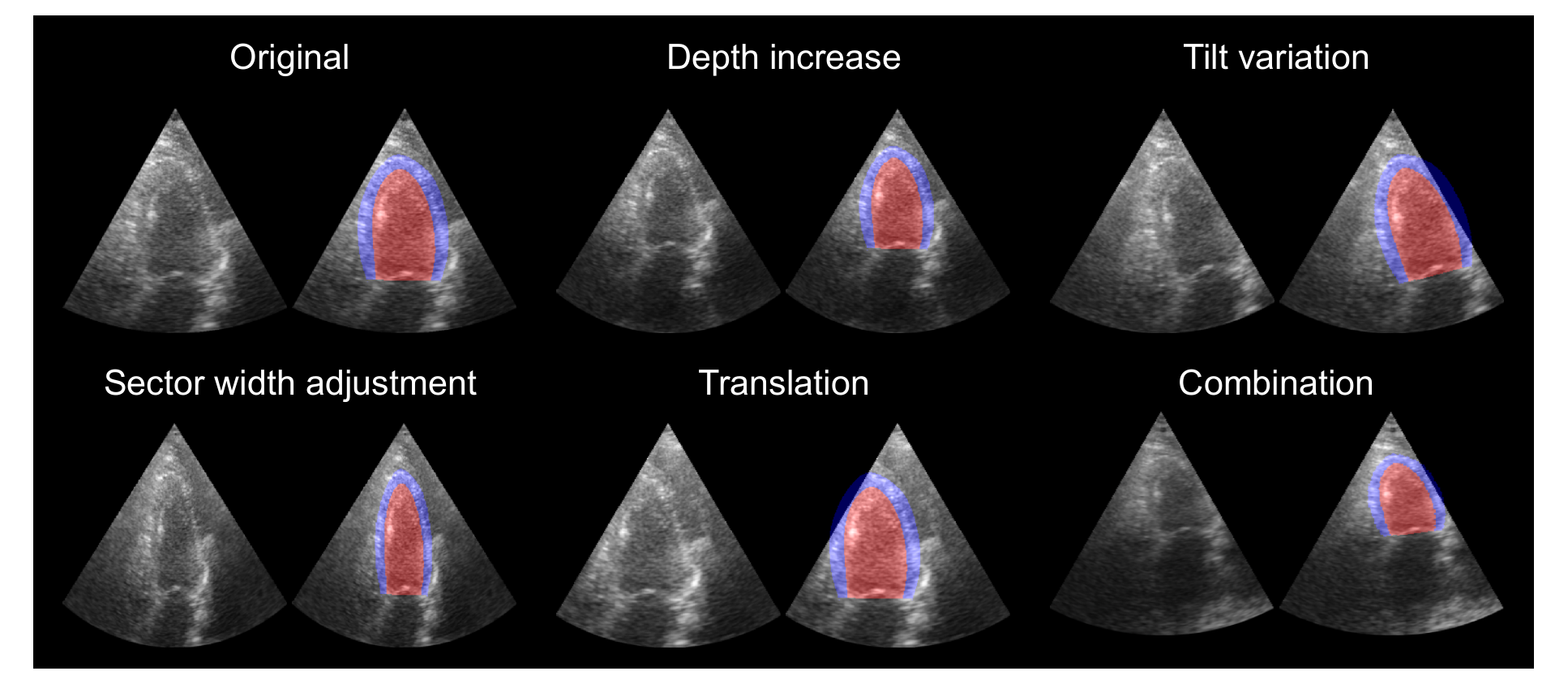}
  \caption{Examples of the generative augmentations types used in this work. All the examples are generated from the same original image shown in the top-left corner.}
  \label{fig: generative_aug_examples}
\end{figure*}

\begin{abstract}

One of the main challenges in current research on segmentation in cardiac ultrasound is the lack of large and varied labeled datasets and the differences in annotation conventions between datasets. This makes it difficult to design robust segmentation models that generalize well to external datasets. 

This work utilizes diffusion models to create generative augmentations that can significantly improve diversity of the dataset and thus the generalisability of segmentation models without the need for more annotated data. The augmentations are applied in addition to regular augmentations. 

A visual test survey showed that experts cannot clearly distinguish between real and fully generated images. 

Using the proposed generative augmentations, segmentation robustness was increased when training on an internal dataset and testing on an external dataset with an improvement of over 20 millimeters in Hausdorff distance. Additionally, the limits of agreement for automatic ejection fraction estimation improved by up to 20\% of absolute ejection fraction value on out of distribution cases.

These improvements come exclusively from the increased variation of the training data using the generative augmentations, without modifying the underlying machine learning model. 

The augmentation tool is available as an open source Python library at
\url{https://github.com/GillesVanDeVyver/EchoGAINS}.
\begin{center}
    \textbf{Keywords}:
\end{center}
Cardiac segmentation, Ultrasound, Generative AI, Diffusion models, RePaint

\end{abstract}

\section{Introduction}

Ischemic heart disease is the leading cause of death worldwide, accounting for 13\% of all global fatalities and is the fastest rising cause of death since the beginning of the century \cite{who_global_health_estimates_2024}. Ultrasound imaging, being cost-effective, safe, and real-time, is the most common technology used for evaluating heart function. Accurately delineating the left ventricle and myocardium directly or indirectly enables the extraction of clinical measurements of heart function, such as ejection fraction (EF) and global longitudinal strain (GLS). However, measurements are labor intensive and even for experienced cardiologists there is high operator-related variability, with a coefficient of variation between 6 and 11\% \cite{armstrong2015quality}. Automating the process of extracting clinical measures from the recordings reduces inter-observer variability and allows measurements over multiple heart cycles, as recommended in the guidelines \cite{lang2015recommendations, olaisen2024automatic, nyberg2024deep, salte2023deep}, without additional labor. \newline

Due to data protection regulations and the labor-intensive process of data curation and annotation, only a limited number of open datasets are available, and those that exist are of limited size. Although guidelines for tracing cardiac structures such as the endocardium exist \cite{lang2015recommendations}, the amount of noise and artifacts in ultrasound, as well as structures such as the trabeculae, make tracing of the true endocardial wall challenging.
Thus, if one would ask multiple experts to annotate the endocardial wall, it would result in multiple different annotations. This variability in annotation preference and inter-vendor differences means that combining datasets from different centers should be done with care. This is in stark contrast with the field of computer vision for natural images, where large, labeled datasets are common. 
The sparsity of annotated data with a consistent annotation protocol creates unique challenges for deep learning networks, particularly in terms of robustness and generalization \cite{chen2020deep}. \newline

Several works have explored using generative AI to improve segmentation in cardiac ultrasound. the idea is to generate images conditioned on segmentation masks to create image-label pairs for training segmentation models. Gilbert et al. \cite{gilbert2021generating} and Tiago et al. \cite{tiago2022data} use generative adversial networks (GANs) for image generation from anatomical models in 2D and 3D respectively. Stojanovski et al. \cite{stojanovski2023echo} developed a conditional Denoising Diffusion Probabilistic
Models (DDPMs) that generates images from segmentation masks. They apply basic transformations to the segmentation masks of an existing labeled dataset and then feed these to the conditional DDPM to generate new images. Jafari et al. \cite{jafari2020cardiac} and Tiago et al. \cite{tiago2023domain} address domain translation, using CycleGANs and adversarial diffusion models respectively. \newline


Generative models trained with conditioning on a segmentation mask have the inherent limitation that they can only be trained on labeled data. In cardiac ultrasound, typically only a portion of the data is labeled with pixel-wise segmentation masks as the annotation process is labor intensive. \newline

In this work, we develop a method that augments a labeled cardiac segmentation dataset using an unconditional diffusion model. Our method has two unique advantages. First, since it uses an unconditionally trained DDPM, a dataset can be augmented using a generative model trained on an unlabeled dataset or a dataset with different annotation conventions. Second, since our model does only alter the surroundings of the segmentation masks, the most crucial parts of the image remain untouched. Thus, fine annotation subtleties and details in the original image-label pair are not affected by the generative model. This distinguishes our approach from the work of Kupyn and Rupprecht \cite{kupyn2025dataset} who repaint the semantic object itself. \newline

The proposed method is most effective for improving the performance of a segmentation model trained on a dataset with limited variation in terms of acquisition and positioning of the left ventricle (LV) in the image. More specifically, this work uses the HUNT4Echo \cite{Olaisen2023-fy} dataset which has annotations of high quality which are time-consuming to obtain. The recordings in this dataset are LV-focused: the recordings were obtained following the clinical guidelines and maximize the area of the ventricle in the image \cite{lang2015recommendations}. Thus, the dataset has limited variety in terms depth, sector width, and positioning of the ventricle in the image. This leads to poor performance of segmentation models trained on this dataset when tested on an external dataset with more variation like the public CAMUS \cite{leclerc2019deep} dataset. In this work, we explore whether generative augmentations can be used to enrich this limited but high-quality dataset so that the resulting models can generalize better to datasets with more variation. \newline

The contributions of this paper are:
\begin{itemize}
    \item A method for creating realistic augmentations of cardiac ultrasound images using DDPMs.
    \item A blinded expert evaluation of the realism of fully generated images.
    \item An ablation study that evaluates the effect of the proposed augmentations on the segmentation accuracy.
    \item A clinical evaluation of the augmentations effect on automatic segmentation-based ejection fraction measurements.
\end{itemize}






\section{Materials and Methods}

\subsection{Datasets}\label{section: Datasets}

The \textbf{HUNT4Echo dataset }\cite{Olaisen2023-fy}, part of the Helse Undersøkelsen i Nord-Trøndelag study, is an ultrasound dataset of 2,211 volunteers of LV-focused apical 2-chamber (A2C) and apical 4-chamber (A4C) views, acquired using a GE Vivid E95 scanner. Each recording includes three cardiac cycles. \newline 

The \textbf{model development set} is a subset that includes single-frame segmentation annotations for both end-diastole (ED) and end-systole (ES), providing pixel-level labels for the left ventricle (LV), left atrium (LA), and myocardium (MYO). The model development set consists of 1058 annotated ED and ES frames of 529 recordings from 311 patients. \newline

The \textbf{ejection fraction set} is a subset disjunct with the model development set of 1900 patients with reference biplane LV volumes in ED and ES. The volumes were obtained following current clinical guidelines by manual tracing and using Simpson's method of discs in the clinically approved EchoPAC software from GE HealthCare on the HUNT4 recordings. \newline 

The \textbf{CAMUS dataset} \cite{leclerc2019deep} is a publicly available dataset containing single cycle recordings from 500 patients, acquired using a GE Vivid E95 scanner (GE Vingmed Ultrasound AS, Norway). The dataset contains one A2C and one A4C recording for each patient and annotations for both the ED and ES frame in each recording, resulting in a total of 2000 image-annotation pairs. Like the HUNT4 dataset, the annotations are pixel-level LV, LA, MYO. The training, validation and test set contain 400, 50 and 50 patients respectively. \newline

\begin{figure}
     \centering
          \begin{subfigure}[b]{0.49\linewidth}
         \centering \includegraphics[trim={0cm 0cm 0cm 0cm}, clip, width=\linewidth]{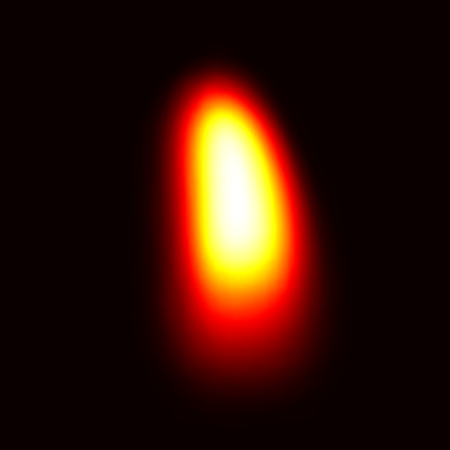}
         \caption{CAMUS \newline}
     \end{subfigure}
     \begin{subfigure}[b]{0.49\linewidth}
         \centering \includegraphics[trim={0cm 0cm 0cm 0cm}, clip, width=\linewidth]{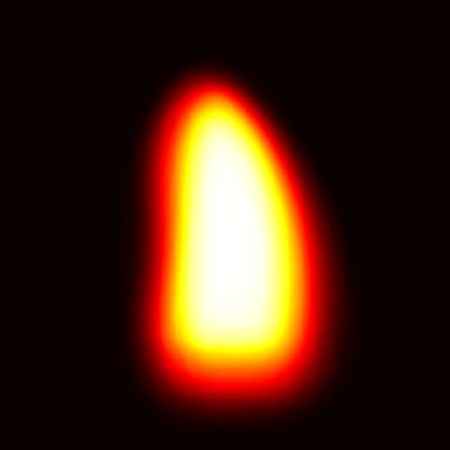}
         \caption{HUNT4 model development set}
     \end{subfigure}
     \begin{subfigure}[b]{0.49\linewidth}
         \centering \includegraphics[trim={0cm 0cm 0cm 0cm}, clip, width=\linewidth]{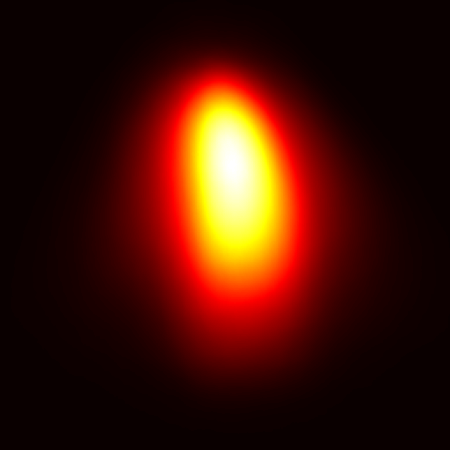}
         \caption{CAMUS with generative augmentations \newline}
     \end{subfigure}
     \begin{subfigure}[b]{0.49\linewidth}
         \centering \includegraphics[trim={0cm 0cm 0cm 0cm}, clip, width=\linewidth]{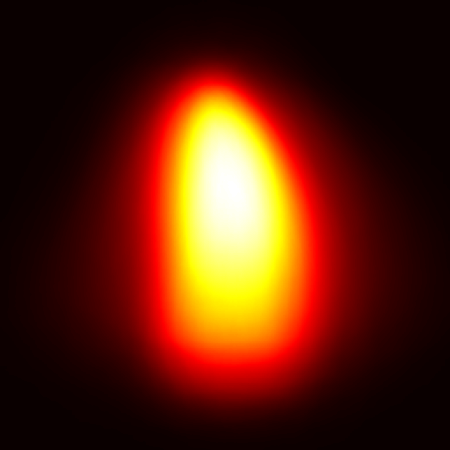}
         \caption{HUNT4 model development set with generative augmentations}
     \end{subfigure}
   \caption{Heatmaps of pixels belonging to the LV after resizing to $256\times256$. This illustrates the difference in scan depth variation and LV positioning in the two datasets, before and after generative augmentations.}
    \label{fig: heatmaps}
\end{figure}

\begin{figure}
     \centering
     \begin{subfigure}[b]{0.49\linewidth}
         \centering \includegraphics[trim={0cm 0cm 0cm 0cm}, clip, width=\linewidth]{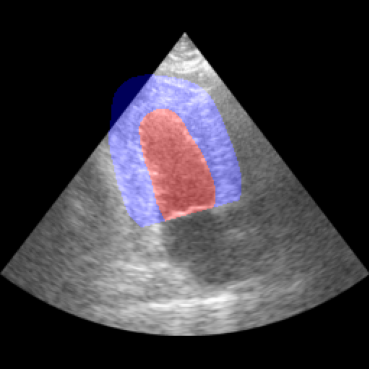}
         \caption{CAMUS}
     \end{subfigure}
     \begin{subfigure}[b]{0.49\linewidth}
         \centering \includegraphics[trim={0cm 0cm 0cm 0cm}, clip, width=\linewidth]{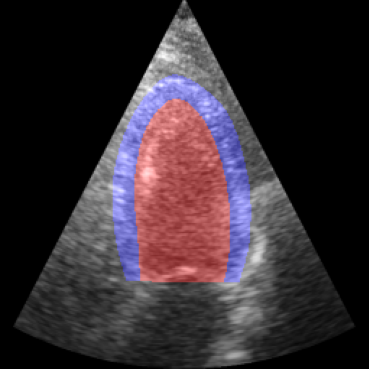}
         \caption{HUNT4}
     \end{subfigure}
   \caption{Example frames from the CAMUS and HUNT4 datasets. The CAMUS and HUNT4 dataset contain the same cardiac views, but the frames in HUNT4 are consistently LV-focused, while those in CAMUS are not. The annotations conventions are also different in both datasets, which can be seen clearly in the thickness of the annotated myocardium (blue).}
    \label{fig: dataset_differences}
\end{figure}

The HUNT4 model development set and the CAMUS dataset should be combined with care due to differences in annotation conventions. For example, the myocardium is consistently annotated as significantly thicker in CAMUS compared to HUNT4. Fig.~\ref{fig: dataset_differences} illustrates this.
There is also a difference in the way the LV is annotated. This is noticeable in the reduced Dice scores in the results when HUNT4 is used for training and CAMUS is used for testing and vica versa, see Table \ref{table: ablation_study_1}).
The segmentation models in this work only segment the LV and MYO labels and the experiments only evaluate on the LV. We elaborate on this choice in the Discussion.
\newline

\begin{figure}[h]
\centering
  \centering
  \includegraphics[trim={0cm 0cm 0cm 0 cm}, clip,width = 1\linewidth]{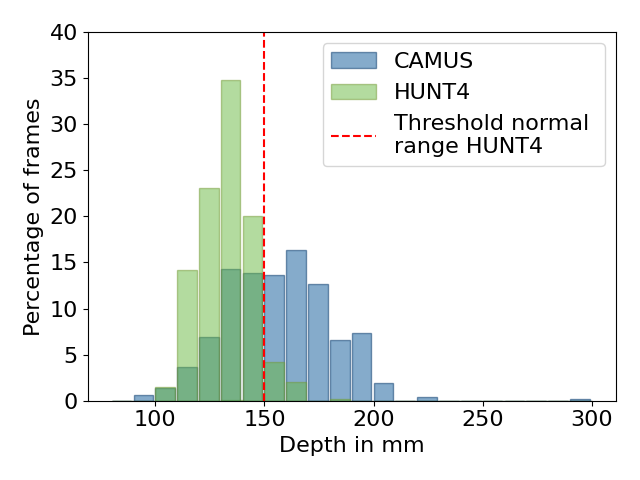}
  \caption{Distribution of imaging depths in CAMUS and HUNT4.}
  \label{fig: depths_hist}
\end{figure}

\begin{figure}[h]
\centering
  \centering
  \includegraphics[trim={0cm 0cm 0cm 0 cm}, clip,width = 1\linewidth]{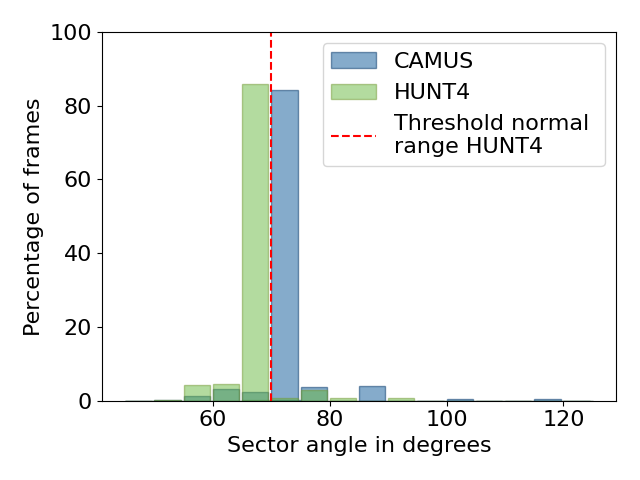}
  \caption{Distribution of sector angles in CAMUS and HUNT4.}
  \label{fig: sector_angles_hist}
\end{figure}

\begin{figure}[h]
\centering
  \centering
  \includegraphics[trim={0cm 0cm 0cm 0 cm}, clip,width = 1\linewidth]{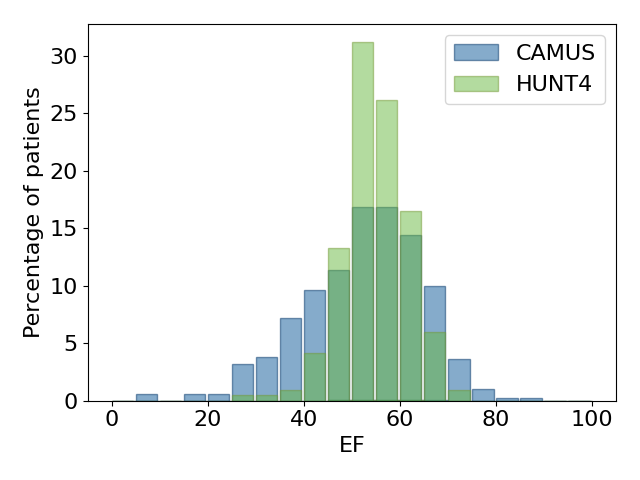}
  \caption{Distribution of EF in CAMUS and HUNT4.}
  \label{fig: ef_hist}
\end{figure}

The recordings in the HUNT4 study follow the clinical guidelines for quantitative measurements and  maximize the area of LV in the image by adjusting the depth \cite{lang2015recommendations}. This is not the case for CAMUS, resulting in less standardized images. The top part of Fig.~ \ref{fig: heatmaps} shows the heatmaps of pixels labeled as LV in both datasets, illustrating that the LV occupies a larger portion of the image in HUNT4 compared to CAMUS. This is a consequence of the distribution of acquisition depth and width, shown in Figs.~\ref{fig: depths_hist} and \ref{fig: sector_angles_hist}. The CAMUS dataset also has a larger variety in terms of EF, as shown in Fig.~\ref{fig: ef_hist}.

\subsection{Denoising Diffusion Probabilistic Models and RePaint} \label{subsection: Denoising Diffusion Probabilistic Models and RePaint}

Denoising Diffusion Probabilistic Models (DDPMs) are a type of generative model that learns to approximate a data distribution by reversing a gradual, multi-step noise addition process. They were first introduced by Sohl-Dickstein et al. \cite{sohl2015deep}, and subsequently improved upon by Ho et al. \cite{ho2020denoising}, and Nichol and Dhariwal \cite{nichol2021improved}. The latter showed that diffusion models can outperform GAN based models for image synthesis \cite{dhariwal2021diffusion}. Diffusion models are also appealing as they do not suffer from the training difficulties often encountered with GANs \cite{esan2023generative}. \newline

The learning methodology of the DDPM has been progressively refined. The DDPM of Ho et al. estimates the noise of the Gaussian target distributions and uses it to calculate the latent of the previous step in the reverse process of the diffusion model. The variance is kept constant with a hyperparameter \cite{ho2020denoising}. Nichol and Dhariwal introduced several improvements to the method of Ho et al. These most important improvements are estimating the variance in addition to the mean, improving the noise scheduler, and using importance sampling during training \cite{nichol2021improved}. This work uses the DDPM described by Nichol and Dhariwal \cite{dhariwal2021diffusion}. \newline

RePaint \cite{lugmayr2022repaint} is a method that use a DDPM to replace specific regions of an image, as controlled by a given mask. During inference, the method takes as input an image and a mask. After each step of the reverse diffusion process, the part of the generated image that falls outside of the mask is replaced by the corresponding parts of the input image, with appropriate noise added for that step in the diffusion process. This ensures that only the image regions inside the mask are synthesized, while those outside are left unchanged, and thus allows the unconditionally trained DDPM to be guided by the input image during inference.

\subsection{Training of the diffusion model}\label{subsection: Training of the DDPM}

\begin{figure}[h]
\centering
  \centering
  \includegraphics[trim={0cm 0cm 0cm 0 cm}, clip,width = 1\linewidth]{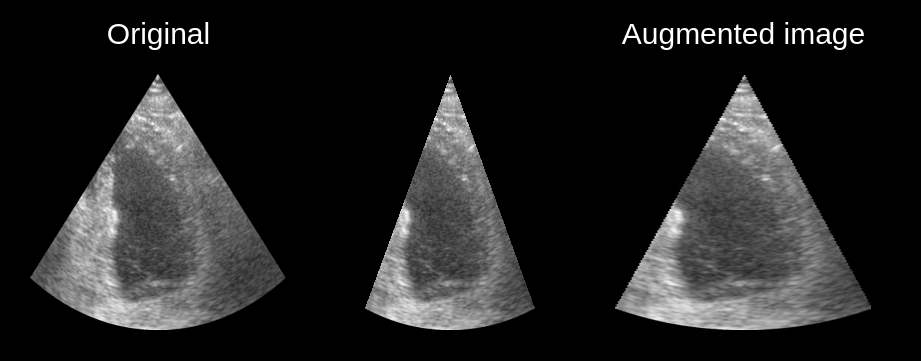}
  \caption{Sector width preprocessing augmentation on CAMUS performed before training of the DDPM. This was done to handle the lack of sector width variation in the CAMUS dataset. The sector angle is first reduced and then the sector is stretched back to 256 by 256 pixels. This step increases the variation in sector widths in the RePaint training set. These augmentations are applied solely during the training of the diffusion model and are not part of the generative augmentations described below.}
  \label{fig: sector_width_aug}
\end{figure}

The goal of the proposed method is to enrich the variety of annotated datasets, so the training dataset of the DDPM should be diverse enough. The CAMUS dataset is mostly diverse enough for this purpose. However, since the majority of sector angles in the acquisitions is around $75^\circ$, a DDPM trained on the CAMUS dataset struggles to generate images with a different sector angle. Therefore, we apply a preprocessing augmentation step that randomly narrows the sector angle by up to 20 degrees, removing pixels along the peripheral scan lines, and then stretches the cut sector back to $256\times256$ pixels, as illustrated in Fig.~\ref{fig: sector_width_aug}. Additionally, to preserve data variety while reducing the training dataset size, only every Nth frame is used, where N is a random number between 8 to 12 frames. This reduces the training set size while preserving the variation since consecutive frames are often very similar due to the high acquisition framerate. \newline


The generative model in this work is the RePaint model as described by Lugmayr et al \cite{lugmayr2022repaint}, where the training and sampling process of the DDPM is replaced by the improved denoising diffusion probabilistic model as described by Nichol and Dhariwal
\cite{nichol2021improved}. Table \ref{table: characteristics diff_unet} summarizes the technical details.

\begin{table}
\scriptsize
  \centering\caption{Key characteristics of the U-Net and its training setup in the diffusion model. The "number of channels" row refers to the number of channels at the first, bottom, and final convolution layers of the U-Net architecture. The "Residual blocks" row refers to the number of blocks per spatial resolution level. For more details, see Nichol and Dhariwal \cite{dhariwal2021diffusion, nichol2021improved}.}
  \renewcommand{\arraystretch}{1} 
  \begin{tabular}{m{86pt}|m{130pt}}
    \toprule
    Number of parameters & 44.1 million \\
    Input size & $256 \times 256$ \\
    Number of channels &  64$\downarrow$ 256 $\uparrow$ 64 \\ 
    Lowest resolution & 8 $\times$ 8\\ 
    Upsampling scheme & Nearest neighbor interpolation\\ 
    Downsampling scheme & Average pooling \\ 
    Normalization scheme & GroupNorm \\ 
    Batch Size & 64 \\
    Optimizer & Adam \\
    Learning rate & 1e-4\\ 
    Learning rate scheduler & None \\ 
    Activation & SiLU \\
    Residual blocks & 3 \\ 
    Training steps & 500k \\
    Self-attention & At resolutions 8 and 16, 4 heads\\
    Diffusion steps & 4000 \\
    Noise scheduler & Cosine \cite{nichol2021improved}\\
    Learn variance & Yes \\
    Loss & Mean squared error, corresponding to the $L_{simple}$ learning objective in \cite{dhariwal2021diffusion}. \\
    \bottomrule
  \end{tabular}
      \label{table: characteristics diff_unet}
\end{table}

\subsection{Generative augmentations} \label{subsection: Generative augmentations}

\begin{figure*}[h]
\centering
  \centering
  \includegraphics[trim={0.75cm 0.25cm 0.75cm 0.25cm}, clip,width = 1\linewidth]{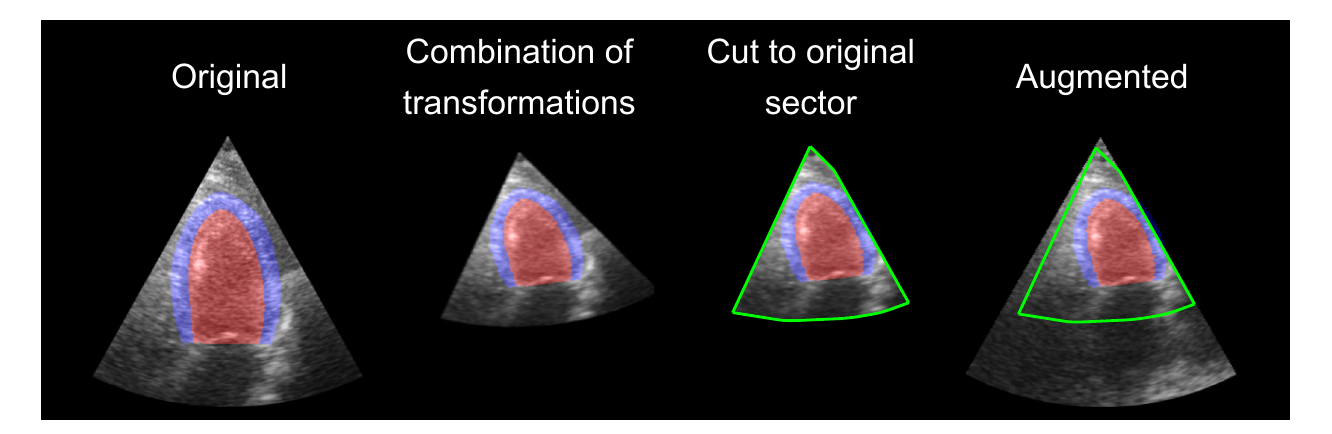}
  \caption{Process of creating generative augmentations. First, the frame is transformed with the transformation described in section \ref{subsection: Generative augmentations}. Then, the pixels outside the original sector are turned black. Finally, the generative model repaints all black pixels re-creating a complete sector in the process. The green contour delineates the part that is kept from the original image.}
  \label{fig: aug_process}
\end{figure*}

To apply generative augmentations to a cardiac ultrasound image, the image is first transformed using random depth, tilt, width and translation transformations as described below and illustrated in Fig.~\ref{fig: generative_aug_examples}. Then the trained diffusion model is applied to synthesize pixels outside of original input image using the RePaint method. Fig.~\ref{fig: aug_process} shows this process. Each image is transformed and augmented five times. The augmented training dataset then contains the original image and the five augmented samples.

\begin{itemize}
    \item \textbf{Depth increase}: the depth of the original image is increases randomly by $\lambda=[0,150]$ pixels by adding black pixels at the bottom of the image and then resizing to $256\times256$.
    \item \textbf{Tilt variation}: the original image is rotated with a random angle around the sector tip by $\theta$ degrees, with $-30^{\circ}<\theta<30^{\circ}$.
    \item \textbf{Sector width adjustment}: the width of the original image is multiplied by a factor $\lambda$. If $\lambda>0$, the image gets stretched out horizontally and cropped back to $256\times256$. If $\lambda<0$, the image gets squeezed into the center. Here, $0.5<\lambda<1.5$. This augmentation is similar to the work of Gazda et al. \cite{gazda2024generative}. Resizing to $256\times256$ distorts the image.
    \item \textbf{Translation}: the original image is shifted by a vector with a random angle and length $\lambda$, with $0<\lambda<50$ pixels. The result is cropped back to $256\times256$ pixels.
    \item \textbf{Combination}: all of the above augmentations are applied with a 50\% chance.
\end{itemize}





\subsection{Survey}

To evaluate the realism of the generated images, a survey was conducted with three groups of human evaluators. The first consisted of three senior cardiologists certified by the EACVI in transthoracic echocardiography, each with over 15 years of experience and more than 10,000 examinations. The second group included four clinical researchers. The last group consisted of three engineers specializing in cardiac ultrasound. Each participant was asked to distinguish real from synthetic images. The real images were sampled randomly from the CAMUS dataset. The synthetic images were generated by the DDPM trained on the CAMUS dataset. The participants were given 50 pairs of images and were told one of the images in each pair was synthetic. The participants than had to select the synthetic image and give an explanation for their selection. Additionally, 5 of the 50 pairs contained two real images without the knowledge of the participants. \newline

\subsection{Segmentation ablation study}

\begin{table}[ht]
\scriptsize
\centering
\caption{Key characteristics of the nnU-Net used for segmentation \cite{isensee2021nnu, nnUNetV2}. The "number of channels" row refers to the number of channels at the first, bottom, and final convolution layers of the U-Net architecture. The "Residual blocks" row refers to the number of blocks per spatial resolution level. The augmentations listed here are performed on top of the proposed generative augmentations. For more details, see Isensee et al. \cite{nnUNetV2}.}
\begin{tabular}{m{83pt}|m{130pt}}
    \toprule
    Number of parameters & 33.4 million \\
    Input size & $256 \times 256$ \\
    Number of channels & 32 $\downarrow$ 512 $\uparrow$ 32 \\ 
    Lowest resolution & $4 \times 4$ \\ 
    Upsampling scheme & Transposed convolutions \\ 
    Downsampling scheme & Strided
convolutions \\ 
    Normalization scheme & InstanceNorm \\ 
    Batch Size & 49 \\
    Optimizer & Adam \\
    Initial learning rate & 1e-2 \\ 
    Learning rate scheduler & Polynomial annealing \\ 
    Loss function & Dice \& cross-entropy \\ 
    Inter-layer activation & Leaky ReLU \\ 
    Final layer activation & Softmax \\ 
    Residual blocks & 2\\
    Epochs & 500 \\
    Deep supervision & At resolutions 128, 64, 32, and 16\\ 
    \\
    Augmentations & Rotations, scaling, Gaussian noise, Gaussian blur, brightness, contrast, simulation of low resolution, gamma correction, and mirroring. \\
    \bottomrule
\end{tabular}
\label{table: characteristics nnunet}
\end{table}

The goal of the ablation study is to evaluate how different types of generative augmentations, described in subsection \ref{subsection: Generative augmentations}, improve segmentation performance. We use the nnU-Net framework \cite{isensee2021nnu, nnUNetV2} as the segmentation model, applying its default configuration but skipping cross-validation. Instead, a single model is trained on the dataset splits defined in section \ref{section: Datasets}. Table \ref{table: characteristics nnunet} lists the key characteristics of the nnU-Net used. \newline

For each type of generative augmentation, we create a training and validation set that combines all the original images frames with five randomly augmented images generated from each original image. Additionally, we use an baseline augmentation that applies the same transformations as the combination augmentation but does not repaint the missing parts, leaving those areas black. This allows us to evaluate whether the repainting actually compared to just applying the transformation augmentations. The resulting sets are then used to train separate models using the nnU-Net framework. The nnU-Net framework applies its regular annotations listed in table \ref{table: characteristics nnunet} on top of the augmentations described here. We evaluate each model’s performance on the original test sets of both CAMUS and HUNT4 datasets.

\subsection{Clinical evaluation on HUNT4}

This experiment compares the automatic segmentation-based ejection fraction (EF) trained with different augmented datasets to the manually measured EF using the clinical EchoPAC software from GE HealthCare. The automated estimation of EF is the same procedure as in previous works \cite{smistad2020real, van2024towards} and follows the steps outlined by the clinical guidelines for manual EF estimation \cite{lang2015recommendations}:

\begin{enumerate}
    \item Use the timing network proposed by Fiorito et al. \cite{fiorito2018detection} to detect the ED and ES frames of each cardiac cycle for both the A2C and A4C recordings of the same patient. Thus, the view is manually labeled during acquisition, while the timing is obtained through deep learning.
    \item Use the segmentation network to segment all ED and ES frames.
    \item Use the modified Simpson method to calculate the LV volume in ED and ES using the A2C and A4C frames. Each A2C frame is combined with each A4C frame for each cardiac cycle and the results are averaged.
    \item Calculate $EF=\frac{{ED\:volume}-ES\:volume}{ED\:volume}$
    
\end{enumerate}

When the segmentation fails for all heart cycles, the algorithm can not extract an EF value and the exam is omitted from analysis, similar to our previous work \cite{van2024towards}. For a fair comparison between the models trained with and without generative augmentations, we only include exams for which both versions manage to extract an EF value. Our results include 1872 out of 1900 patients in the HUNT4 EF set, which corresponds to a feasibility of 98.5\%. \newline

\subsection{Real-time demo}

\begin{figure}[h]
\centering
  \centering
  \includegraphics[trim={0.2cm 0.2cm 0.2cm 0.2cm}, clip,width = 1\linewidth]{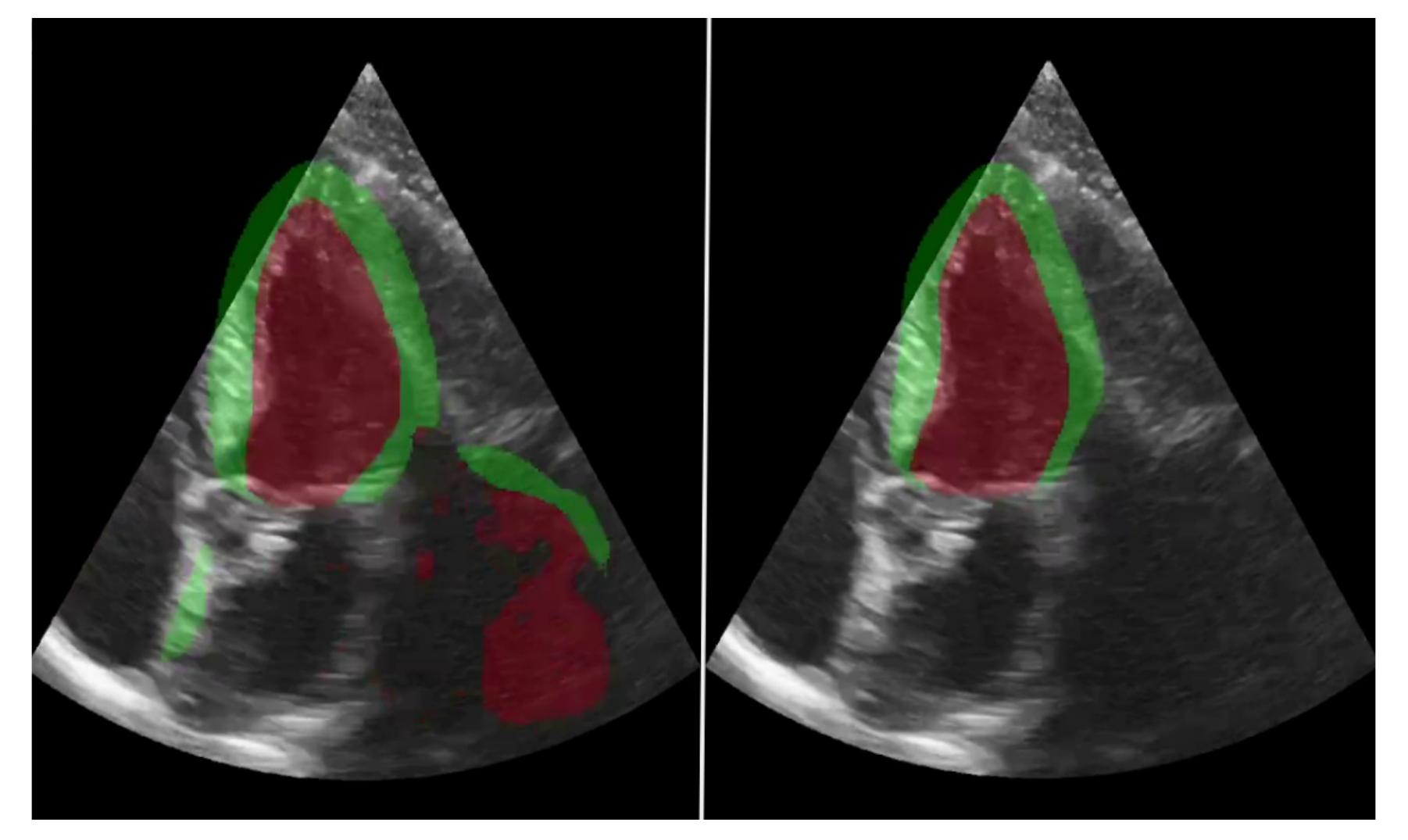}
  \caption{Screenshot of the real-time demo application. The left side shows the segmentation of the segmentation model trained in the usual way. The right side shows the segmentation of the same model trained with the combination of all generative augmentations.}
  \label{fig: real_time_demo}
\end{figure}

To visually demonstrate the differences between the model trained with and without generative augmentations with varying acquisition parameters, a real-time application was created using the FAST framework \cite{smistad2015fast}. The application shows the segmentation output of the segmentation model trained without generative augmentations and the model trained with the combination of all generative augmentations side by side in real-time while streaming from a GE HealthCare Vivid E95 scanner. Fig.~\ref{fig: real_time_demo} shows a screenshot of the application. The video is available at \url{https://doi.org/10.6084/m9.figshare.28219919}.
These clearly demonstrate an increased segmentation model robustness in terms of acquisition parameters such as depth, angle and LV positioning.

\begin{figure*}
\centering
  \centering
  \includegraphics[trim={0cm 0cm 0cm 0 cm}, clip,width = 0.618\linewidth]{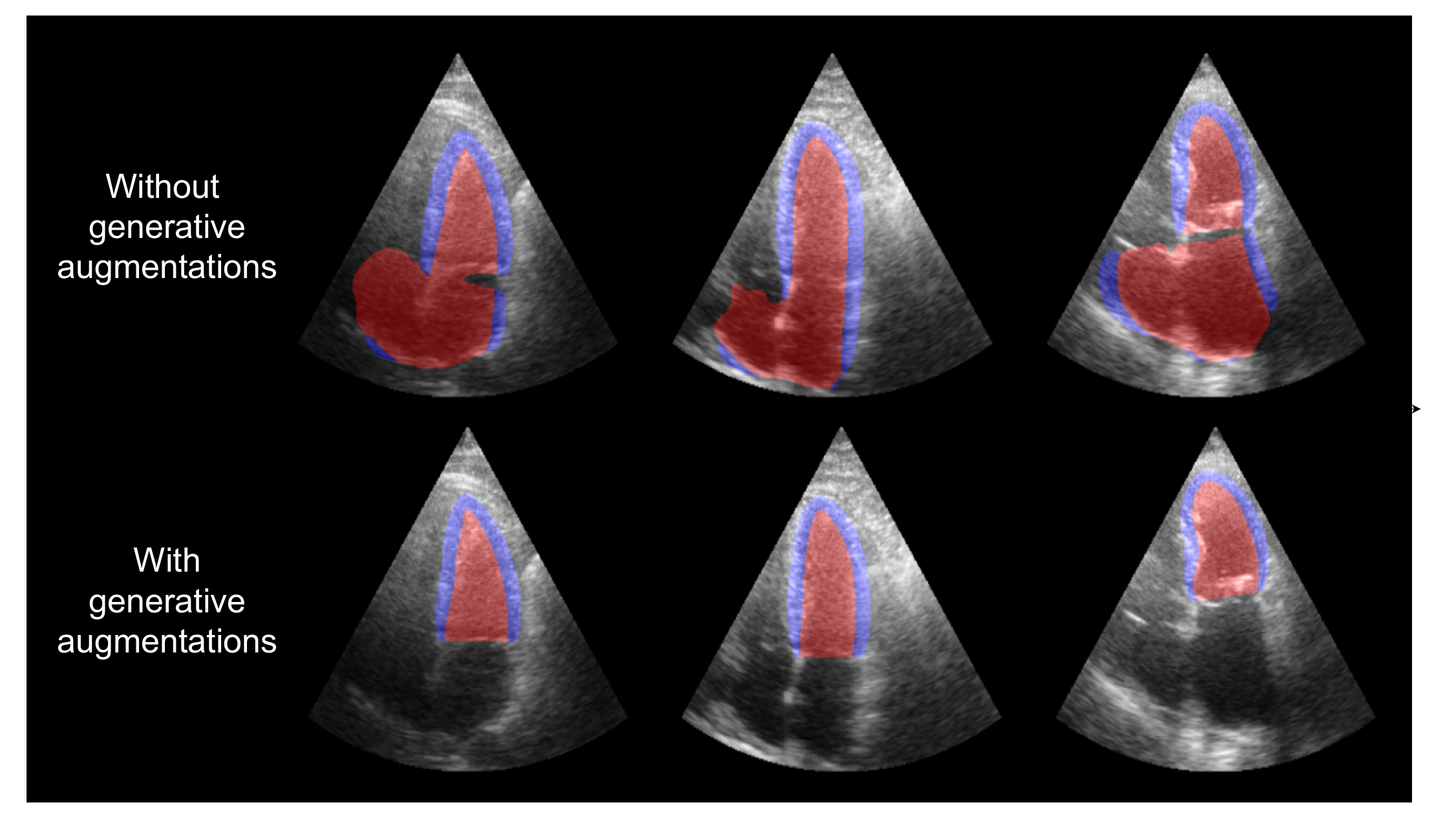}
  \caption{Segmentation results for HUNT4 study participants with the largest difference in automatic EF for models with and without generative augmentations. The model trained without generative augmentations fails to correctly segment the LV  for frames with increased depth due to the lack of such images in the training set.}
  \label{fig: biggest_diff.pdf}
\end{figure*}

\begin{figure}
\centering
  \centering
  \includegraphics[trim={0cm 0cm 0cm 0 cm}, clip,width = 0.5\linewidth]{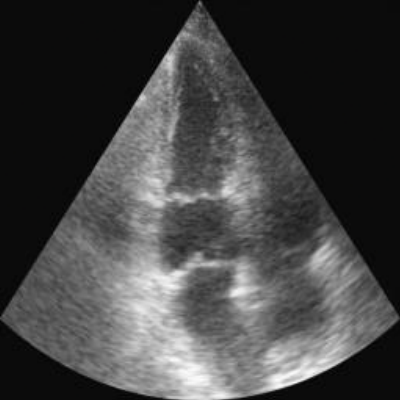}
  \caption{A problematic hallucination by the diffusion model, generating a second mitral valve below the true mitral valve.}
  \label{fig: hallucination_repaint}
\end{figure}

\section{Results}

\subsection{Evaluation of generated images} \label{subsection: Evaluation of generated images}

\begin{figure*}[h]
\centering
  \centering
  \includegraphics[trim={0.75cm 0.25cm 0.75cm 0.25cm}, clip,width = 1\linewidth]{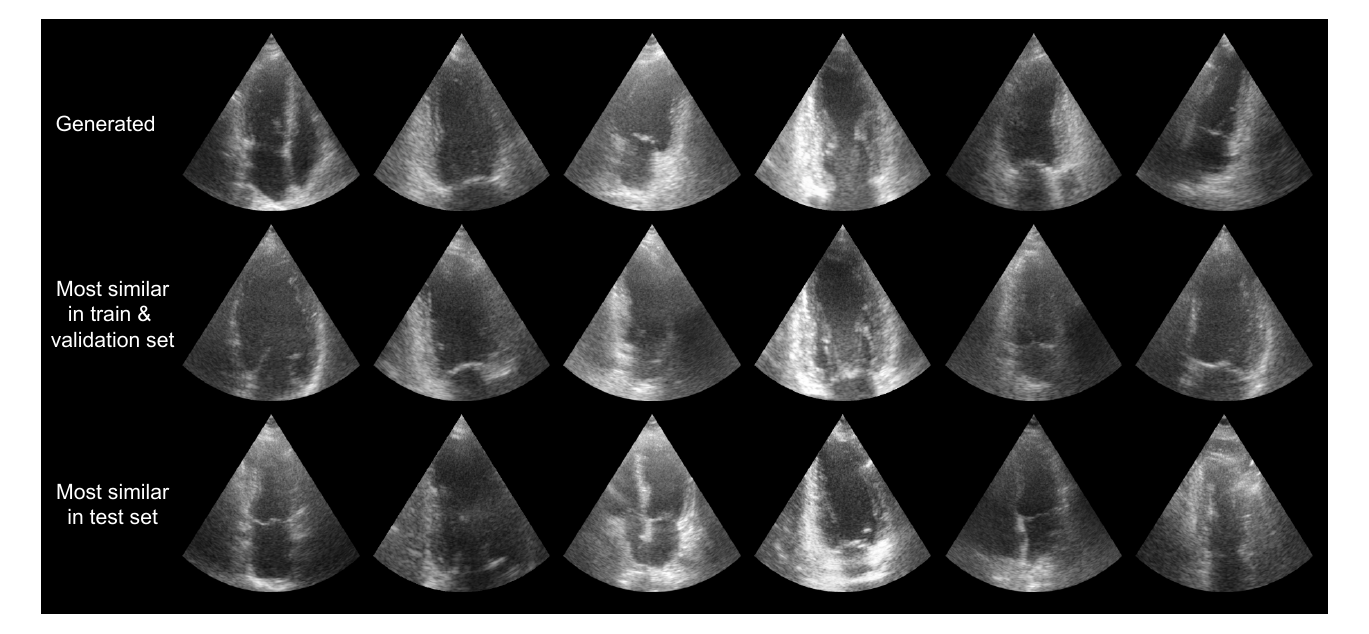}
  \caption{Generated samples, together with most similar cases in the train and validation set and the test set of the CAMUS dataset, based on SSIM \cite{wang2004image}.}
  \label{fig: similar_samples}
\end{figure*}

The ImageNet Fréchet inception distance (FID) \cite{heusel2017gans} and inception score (IS) \cite{salimans2016improved}
of the diffusion model are 23.87 and 1.47 respectively. However, these metrics can give misleading results for generative models that are not trained on ImageNet \cite{deng2009imagenet, barratt2018note, rosca2017variational}. To qualitatively assess the performance of the model, Fig.~\ref{fig: similar_samples} shows random samples generated together with the most similar cases from the CAMUS dataset identified automatically using the structural similarity index measure (SSIM) \cite{wang2004image}. This shows the model does not simply memorize cases from the training set, and produces realistic and varied samples. \newline

\subsection{Survey results}

\begin{figure*}[h]
\centering
  \centering
  \includegraphics[trim={0cm 0cm 0cm 0cm}, clip,width = 1\linewidth]{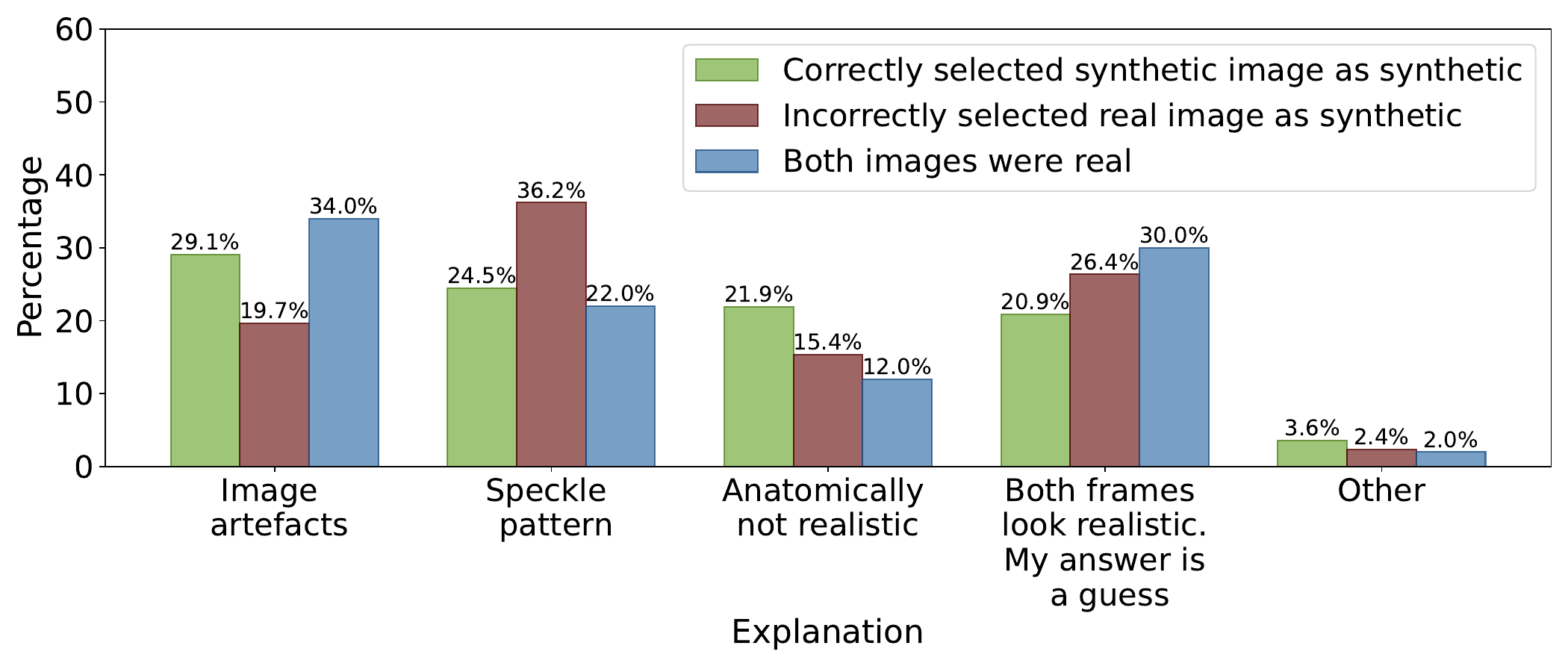}
  \caption{Explanations given during the survey}
  \label{fig: survey}
\end{figure*}

On the 45 pairs with one real and one synthetic image, participants correctly identified the synthetic image 56.4\% of the time. When broken down by group, cardiologists achieved an accuracy of 63.7\%, while clinical researchers and engineers both identified the correct frame 53.3\% of the time. Fig.~\ref{fig: survey} shows the explanations given when the participants correctly identified the synthetic frame, when they were wrong, and when both frames were real in the 5 cases mentioned above.
\newline

Using a binomial test with a significance level of 5\%, the accuracy of the cardiologists was found to be statistically significantly higher than random guessing ($P=0.09\%$). However, the engineers and clinical researchers in the survey did not show statistically significant higher accuracy compared to random guessing ($P=24.6\%$).

\subsection{Segmentation ablation study results}

Table \ref{table: ablation_study_1} shows the results of the ablation study on the CAMUS dataset, using Dice score and Hausdorff distance as metrics. The bottom part of Fig.~\ref{fig: heatmaps} shows the heatmaps of pixels belonging to the LV after applying the combination of all generative augmentations. Comparing these to the original illustrates that the generative augmentations increase the variety of LV location in the image. \newline

The increase in segmentation accuracy of the HUNT4 model on CAMUS originate mostly from an improvement in segmentation accuracy for samples outside the HUNT4 image distribution. Table \ref{table: camus_subsets_results} lists the segmentation results for the HUNT4 models on different subsets of CAMUS. The subsets are based on depth and sector angle cutoff values visualized in Figs.~\ref{fig: depths_hist} and \ref{fig: sector_angles_hist}.


\begin{table*}[h]
\scriptsize
  \centering
  \renewcommand{\arraystretch}{1} 
  \caption{Segmentation results of the ablation study using different datasets (HUNT4 and CAMUS) for training and testing. For all experiments, regular augmentations are applied in addition to the generative augmentations (see Table \ref{table: characteristics nnunet}).The Dice score and Hausdorff distance are only for the LV lumen label. We elaborate on this choice in the Discussion. Since the two datasets have been annotated by different experts with different annotation conventions, there is a considerably lower segmentation accuracy when the training and test sets are different. }
  \begin{tabular}{m{60pt}m{40pt}m{120pt}m{60pt}m{100pt}}
    \toprule
      Training set  & Test set & Generative Augmentations & Dice score & Hausdorff distance (mm)\\
    \midrule
    \multirow{7}{1.4cm}{HUNT4} & \multirow{7}{1.4cm}{CAMUS} & None & 0.802 $\pm$ 0.15 & 29.03 $\pm$ 26.01\\
    && Depth increase & 0.887 $\pm$ 0.05 & \textbf{7.49 $\pm$ 3.25} \\
    && Tilt variation  & 0.829 $\pm$ 0.14 & 17.31 $\pm$ 20.98 \\
    && Sector width & 0.847 $\pm$ 0.11 & 21.36 $\pm$ 23.84\\
    && Translation & 0.840 $\pm$ 0.12 & 16.55 $\pm$ 19.71 \\
    && Combination & \textbf{0.887 $\pm$ 0.05} & 8.17 $\pm$ 5.32 \\
    && Combination without repaint & 0.810 $\pm$ 0.15  & 26.90 $\pm$ 25.07\\
    \midrule
    \multirow{7}{1.4cm}{CAMUS} &  \multirow{7}{1.4cm}{CAMUS} & None & 0.943 $\pm$ 0.03 & 4.46 $\pm$ 2.52 \\
    && Depth increase & 0.945 $\pm$ 0.03 & \textbf{4.27 $\pm$ 2.34} \\
    && Tilt variation  &  0.945 $\pm$ 0.03 &  4.30 $\pm$ 2.43 \\
    && Sector width variation & \textbf{0.946 $\pm$ 0.03} & 4.34 $\pm$ 2.41 \\
    && Translation & 0.944 $\pm$ 0.03 & 4.44 $\pm$ 2.43\\
    && Combination & 0.944 $\pm$ 0.03 & 4.37 $\pm$ 2.43 \\
    && Combination without repaint & 0.934 $\pm$ 0.03 & 5.39 $\pm$ 2.85 \\

      \midrule
      \midrule
        \multirow{7}{1.4cm}{HUNT4} & \multirow{7}{1.4cm}{HUNT4} & None &  0.952 $\pm$ 0.02 &  3.34 $\pm$ 1.21 \\
    && Depth increase & 0.954 $\pm$ 0.02 & 3.24 $\pm$ 0.99 \\
    && Tilt variation & 0.954 $\pm$ 0.02 & 3.38 $\pm$ 1.06 \\
    && Sector width variation & 0.953 $\pm$ 0.02  & \textbf{3.23 $\pm$ 1.00} \\
    && Translation & 0.954 $\pm$  0.02  & 3.32 $\pm$  0.97 \\
    && Combination & \textbf{0.954 $\pm$ 0.02} & 3.31 $\pm$ 0.99 \\
    && Combination without repaint &  0.947 $\pm$ 0.02 & 4.14 $\pm$ 1.85 \\
    \midrule
    \multirow{7}{1.4cm}{CAMUS} &\multirow{7}{1.4cm}{HUNT4} & None & 0.886 $\pm$ 0.04 & 6.70 $\pm$ 1.81 \\
    && Depth increase & 0.891 $\pm$ 0.04 & 6.55 $\pm$ 1.84 \\
    && Tilt variation  & 0.887 $\pm$ 0.04 & 6.69 $\pm$ 1.91 \\
    && Sector width variation &  0.892 $\pm$ 0.04 & \textbf{6.54 $\pm$ 1.78} \\
    && Translation &  0.890 $\pm$ 0.04  & 6.55 $\pm$ 1.83   \\
    && Combination & \textbf{0.892 $\pm$ 0.04} & 6.59 $\pm$ 1.82 \\
    && Combination without repaint & 0.875 $\pm$ 0.04 & 7.71 $\pm$ 2.11 \\
    \bottomrule
  \end{tabular}
      \label{table: ablation_study_1}
\end{table*}

\begin{table*}
\scriptsize
  \centering
  \renewcommand{\arraystretch}{1} 
  \caption{Segmentation results on different CAMUS subsets for a segmentation model trained on HUNT4 without generative augmentations and with the combination of all
  generative augmentations. }
  \begin{tabular}{m{100pt}m{120pt}m{100pt}m{100pt}}
    \toprule
      Training dataset   & CAMUS Test subset & Dice score & Hausdorff distance (mm)\\
        \midrule
      \multirow{4}{4cm}{HUNT4 without generative augmentations} & Depth $< 150$ mm ($n=1088$) &   0.855 $\pm$ 0.11  & 14.48 $\pm$ 16.61 \\
          &  Depth $\geq 150$ mm ($n=912$) & 0.729 $\pm$ 0.18 & 45.83 $\pm$ 30.19 \\
      &  Sector angle $< 70^\circ$ ($n=146$)& 0.869 $\pm$ 0.10 & 12.47 $\pm$ 16.47 \\
      &   Sector angle $\geq 70^\circ$ ($n=1854$) & 0.792 $\pm$ 0.16  & 30.06 $\pm$ 28.80 \\
      \midrule
      \multirow{4}{4cm}{HUNT4 with generative augmentations} & Depth $< 150$ mm ($n=1088$) &  \textbf{0.893 $\pm$ 0.05} & \textbf{7.45 $\pm$ 3.80}   \\
      &  Depth $\geq 150$ mm ($n=912$) & \textbf{0.886 $\pm$ 0.07} & \textbf{9.34 $\pm$ 8.37} \\
      &  Sector angle $< 70^\circ$ ($n=146$)& \textbf{0.893 $\pm$ 0.05}  & \textbf{7.11 $\pm$ 3.10} \\
      &   Sector angle $\geq 70^\circ$ ($n=1854$) &  \textbf{0.890 $\pm$ 0.07} & \textbf{8.40 $\pm$ 6.56} \\
    \bottomrule
  \end{tabular}
      \label{table: camus_subsets_results}
\end{table*}


\subsection{Clinical evaluation on HUNT4 results}

\begin{figure*}[h!]
    \centering
    \begin{subfigure}{0.382\linewidth}
        \centering
        \includegraphics[trim={1cm 1.6cm 0.4cm 2cm},clip,width=\textwidth]{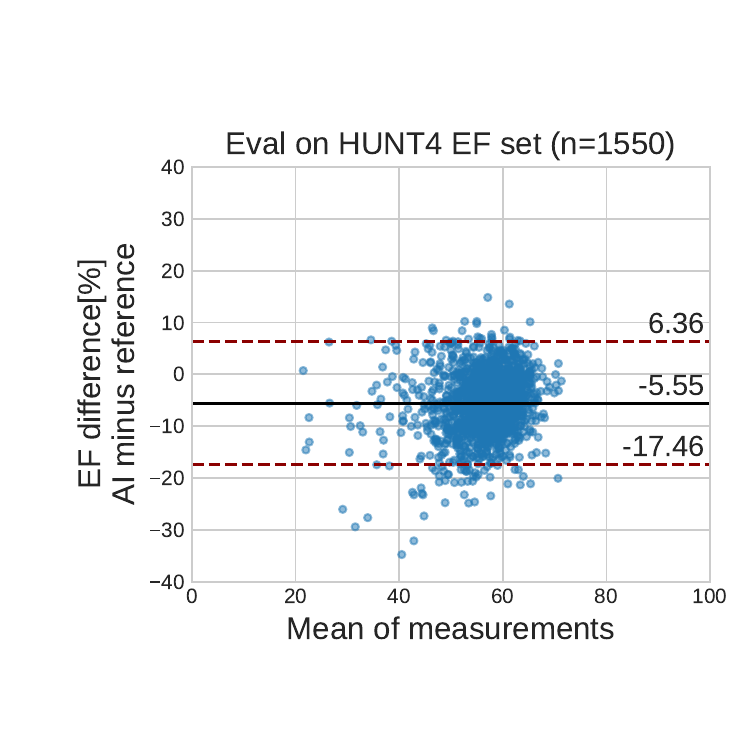}
        \caption{Trained on \textbf{HUNT4} \textbf{without} generative augmentations, tested \textbf{in normal range}.}
    \end{subfigure}
    \hspace{1.5cm}
    \begin{subfigure}{0.382\linewidth}
        \centering
        \includegraphics[trim={1cm 1.6cm 0.4cm 2cm},clip,width=\textwidth]{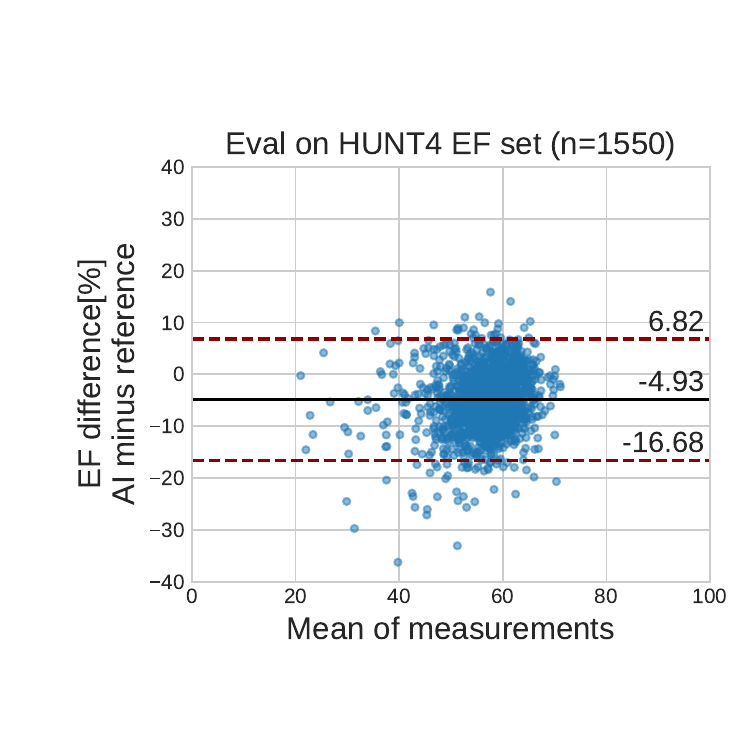}
        \caption{Trained on \textbf{HUNT4} \textbf{with} generative augmentations, tested \textbf{in normal range}.}
    \end{subfigure}


    \begin{subfigure}{0.382\linewidth}
        \centering
        \includegraphics[trim={1cm 1.6cm 0.4cm 2cm},clip,width=\textwidth]{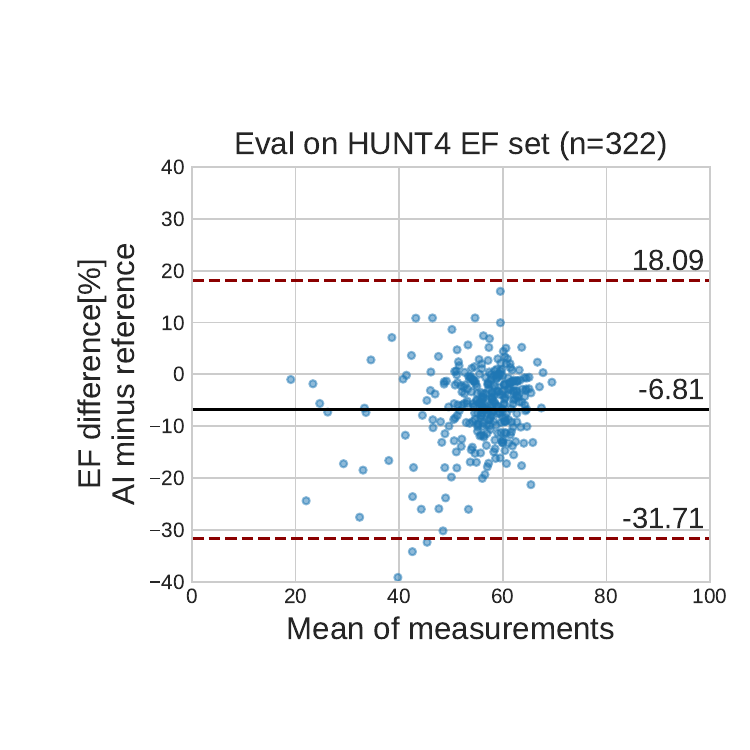}
        \caption{Trained on \textbf{HUNT4} \textbf{without} generative augmentations, tested \textbf{outside normal range}.}
    \end{subfigure}
    \hspace{1.5cm}
    \begin{subfigure}{0.382\linewidth}
        \centering
        \includegraphics[trim={1cm 1.6cm 0.4cm 2cm},clip,width=\textwidth]{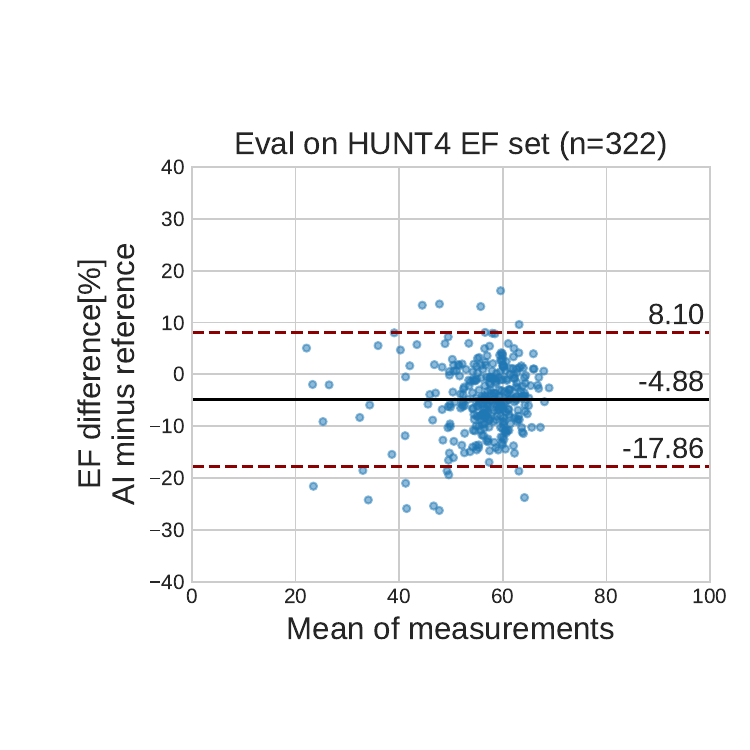}
        \caption{Trained on \textbf{HUNT4} \textbf{with} generative augmentations, tested \textbf{outside normal range}. \newline}
    \end{subfigure}


    \begin{subfigure}{0.382\linewidth}
        \centering
        \includegraphics[trim={1cm 1.6cm 0.4cm 2cm},clip,width=\textwidth]{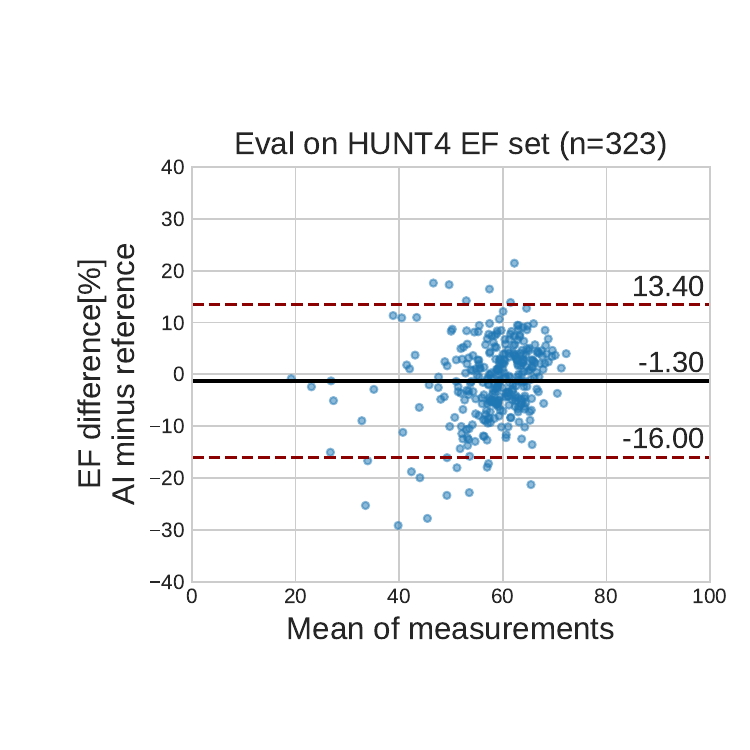}
        \caption{Trained on \textbf{CAMUS} \textbf{without} generative augmentations, tested \textbf{outside normal range}.}
    \end{subfigure}
    \hspace{1.5cm}
    \begin{subfigure}{0.382\linewidth}
        \centering
        \includegraphics[trim={1cm 1.6cm 0.4cm 2cm},clip,width=\textwidth]{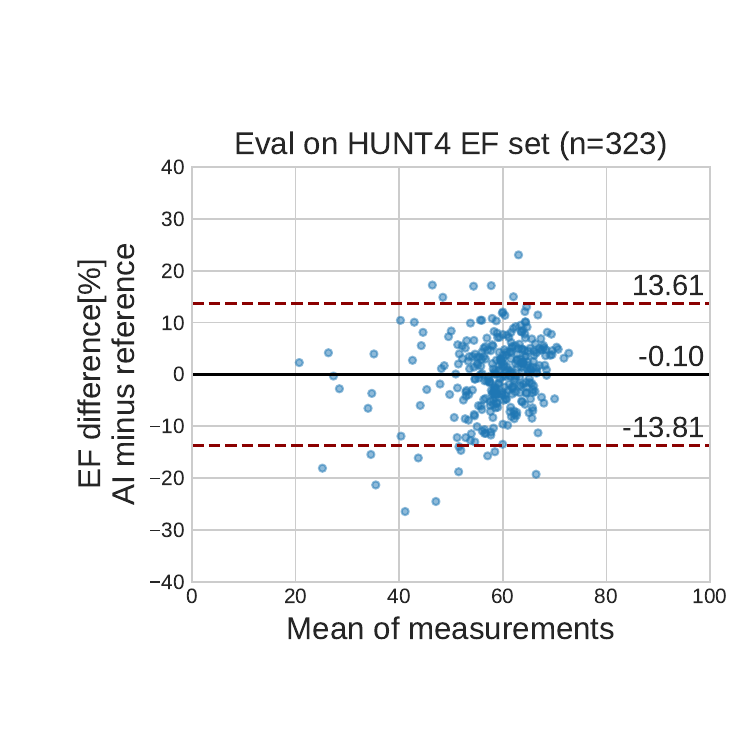}
        \caption{Trained on \textbf{CAMUS} \textbf{with} generative augmentations, tested \textbf{outside normal range}. \newline}
    \end{subfigure}

    \caption{Bland–Altman plots comparing the manual reference with automatic EF measurements obtained via segmentation trained with and without generative augmentations. The exams outside the normal range are the exams where at least one frame used in the calculation is outside the normal range of HUNT4 (depth $>150$mm or sector angle $>70^\circ$).}
    \label{fig: ef_main_text}
\end{figure*}

Similar to the segmentation results, the performance gains of the HUNT4 model originate mostly from an improvement in segmentation accuracy for frames outside the normal range. Fig.~\ref{fig: ef_main_text} shows the Bland-Altman plots comparing the manual reference EF with the automatic EF for segmentation models trained with and without generative augmentations for data both inside and outside of the HUNT4 acquisition normal range of depth $> 150$mm and sector angle $> 70^\circ$. Appendix \ref{appendix: exensive EF evaluation} contains additional analysis of automatic EF and also evaluates automatic on CAMUS.



\section{Discussion}

The results of the survey shows that the DDPM can generate highly realistic ultrasound images that are hard to distinguish, even for ultrasound experts. Although senior cardiologists could distinguish synthetic images better than random guessing, they were still correct only in 63 7\% of the cases. \newline

Generative AI can create unrealistic and anatomically incorrect images, also known as hallucinations, and the DDPM in this work is no exception. Fig.~\ref{fig: hallucination_repaint} shows an example of a problematic hallucination in which the model creates an additional mitral valve. In this case, the augmentation would add noise to the training data. For segmentation augmentations, the most crucial parts are the parts with the reference segmentation masks, which are original and real. The remaining background region is of less importance and in most cases it is not detrimental if the generated surroundings are not perfectly accurate, although there are exceptions as the example in Fig.~\ref{fig: hallucination_repaint}. Still, the comparison to the baseline augmentations where the surrounding pixels remain black shows that it is still important that the surroundings look realistic. \newline

Our study does not include segmentation of the LA because it is often not fully visible in the original scan. This is especially true for the LV-focused HUNT4 images. When using data augmentation, particularly depth augmentation, the diffusion model can generate parts of the LA that were missing in the original scan. This creates problems because the original labels only correspond to the visible parts of the LA in the original image. Therefore, we restrict ourselves to the LV and myocardium (MYO) in this work. \newline

While the segmentation models predict both LV and MYO, the experiments only evaluate on the LV. The experiments do not evaluate on the MYO because there is a notable difference between the annotation conventions between HUNT4 and CAMUS. The annotations in the CAMUS dataset consistently label the MYO notably thicker than the HUNT4. Fig.~\ref{fig: dataset_differences} shows an example of this. There are also differences in annotation conventions for the LV lumen, but these are less pronounced than for the MYO. \newline

\noindent
The clinical evaluation on HUNT4 showed the generative augmentations lead to narrower limits of agreement with the reference in terms of EF. The reduction in limits of agreement originates from a reduction in segmentation failures (outliers). Fig. \ref{fig: biggest_diff.pdf} shows visualizations of segmentation outputs for HUNT4 study participants where the segmentation models with and without augmentations lead to the largest differences in automatic EF. In the training set of the HUNT4 development set, all views are LV-focused meaning that a shorter scan depth is used so that the LV covers most of the scan sector. LV-focused views are used because this aligns with clinical guidelines, which recommend optimizing the view to ensure the left ventricle is clearly and fully visualized for accurate assessment. However, in practice it can be hard to get a standardized view. Without the generative augmentations, the model overfits on LV-focused views and thus often fails to segment the LV correctly when views are not focused on the LV. This explains why the depth augmentation are the most successful augmentation in the ablation study. \newline

The clinical evaluation shows that the bias changes depending on which dataset the segmentation model was trained on. This bias can be corrected for by measuring its magnitude on a subset of the target domain and adjusting accordingly on new, unseen data. The reason for the change of bias can be the data distribution in the training set, the annotation conventions, or the methodology of the tools used for manual measurement \cite{olaisen2024automatic}. An in-depth analysis of the bias is out of scope for this work. \newline

Both the ablation study and the clinical evaluation on HUNT4 show that the model trained on HUNT4 benefits the most from the generative augmentations. 
The HUNT4 dataset is more standardized, contains recordings from mostly healthy volunteers and contains less variation than the CAMUS dataset. Thus, the increase in variation from the generative augmentations mostly benefits this dataset. However, also the segmentation model trained and tested on CAMUS shows small, but statistically significant ($p<0.05$), improvement using the Wilcoxon signed-rank test \cite{woolson2007wilcoxon}. \newline

The proposed generative augmentations improve the variation in terms of acquisition, positioning and size of the LV in the image, but do not diversify in terms of shape of the heart itself. In practice, these two types of variation might be correlated, as more diversity in patients would naturally lead to more variation in scan sectors. \newline



The study only looks into generative augmentations for segmentation on cardiac images, but there is nothing preventing similar generative augmentation methods to be applied to other tasks or other imaging modalities. Of course, the type of generative augmentations in this work are tailored to cardiac ultrasound, but the concept of generative augmentations itself is flexible. Any segmentation task for which enriching the positioning of the reference masks can not be achieved realistically with regular augmentations could use the proposed approach. \newline

\section{Conclusion}

This work explores using generative augmentations for cardiac ultrasound segmentation. Our results show that diffusion models can generate highly realistic cardiac ultrasound images indistinguishable from real images by experts. We show how these generative models can be used to improve segmentation model accuracy and generalizability through generative augmentations. \newline

The proposed generative augmentations are most useful for datasets with limited variation in terns of acquisition and positioning of the left ventricle in the image. This is relevant in the medical domain, as datasets are often limited according to the acquisition protocol and preference of the personnel at a given center. \newline

Although this work only studies segmentation for cardiac ultrasound, the concept of generative augmentations could be generalized to other tasks or imaging modalities.


\FloatBarrier

\bibliographystyle{IEEEtran}

\bibliography{bibliography.bib}

\onecolumn

\appendix

\section{Extensive evaluation of automatic EF} \label{appendix: exensive EF evaluation}

\begin{figure*}[h]
    \centering
    \begin{subfigure}{0.382\linewidth}
        \centering
        \includegraphics[trim={1cm 1.6cm 0.4cm 2cm},clip,width=\textwidth]{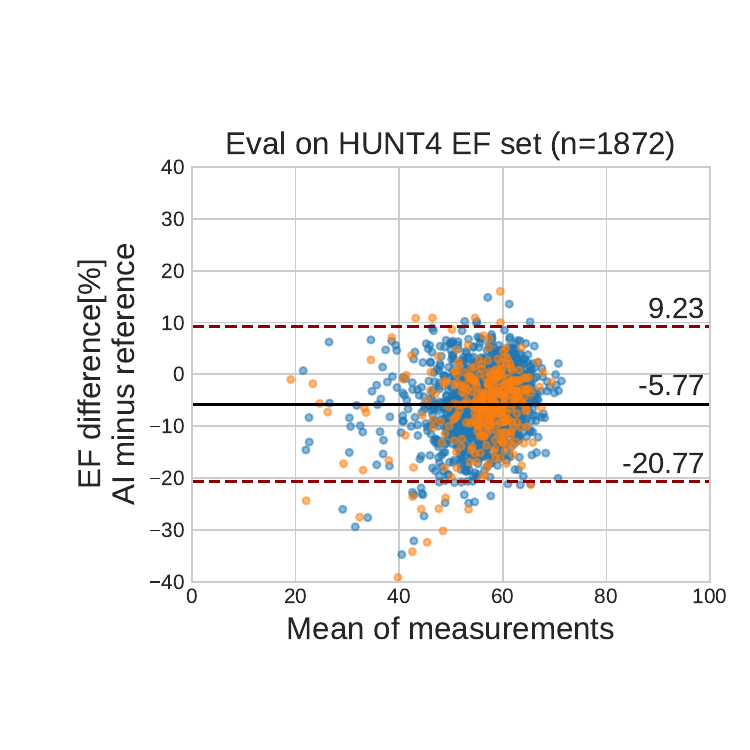}
        \caption{Trained on \textbf{HUNT4} \textbf{without} generative augmentations}
    \end{subfigure}
    \hspace{1.5cm}
    \begin{subfigure}{0.382\linewidth}
        \centering
        \includegraphics[trim={1cm 1.6cm 0.4cm 2cm},clip,width=\textwidth]{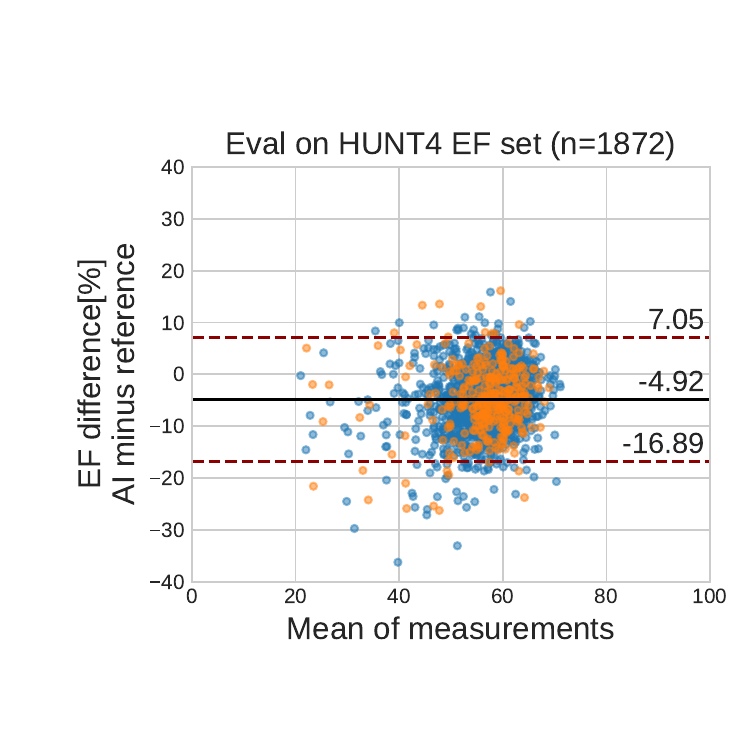}
        \caption{Trained on \textbf{HUNT4} \textbf{with} generative augmentations}
    \end{subfigure}


    \begin{subfigure}{0.382\linewidth}
        \centering
        \includegraphics[trim={1cm 1.6cm 0.4cm 2cm},clip,width=\textwidth]{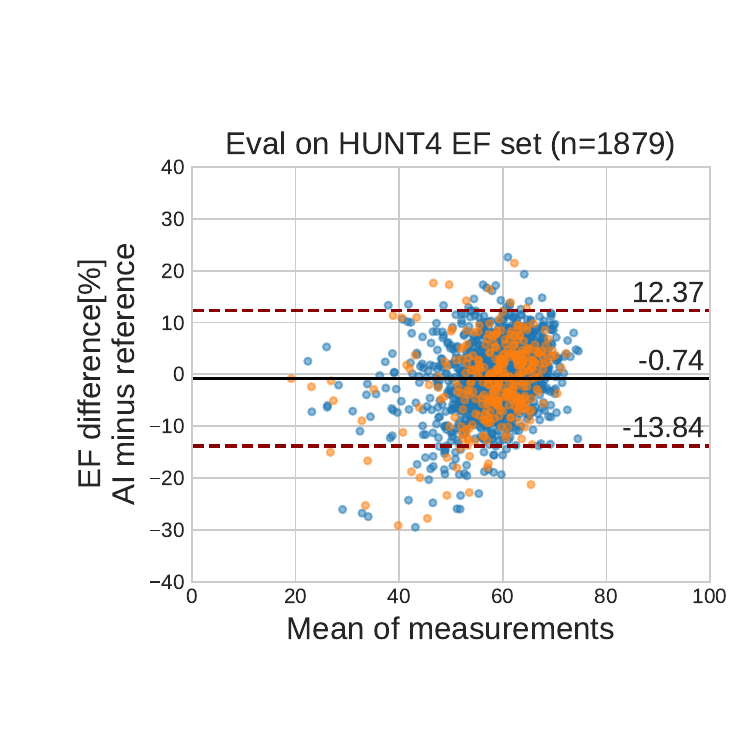}
        \caption{Trained on \textbf{CAMUS} \textbf{without} generative augmentations}
    \end{subfigure}
    \hspace{1.5cm}
    \begin{subfigure}{0.382\linewidth}
        \centering
        \includegraphics[trim={1cm 1.6cm 0.4cm 2cm},clip,width=\textwidth]{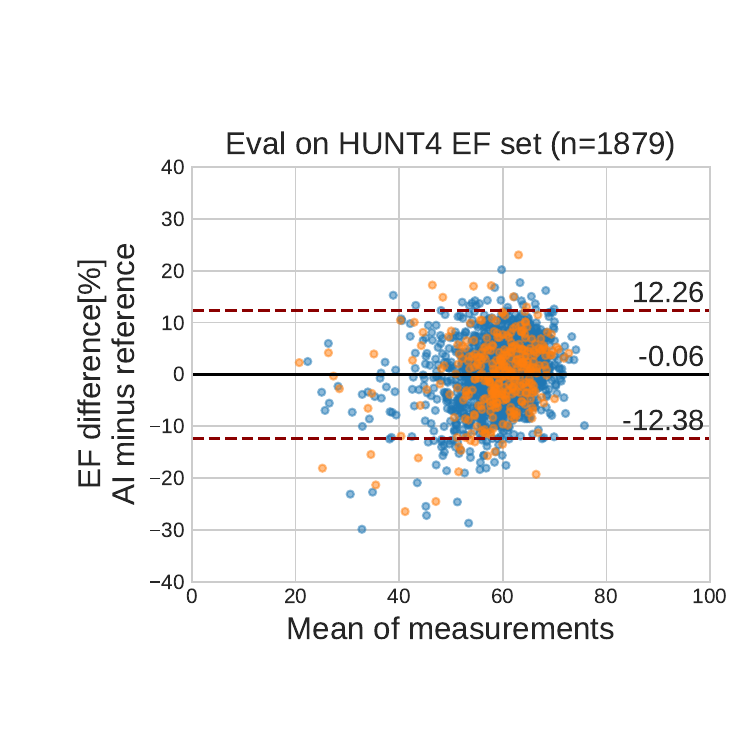}
        \caption{Trained on \textbf{CAMUS} \textbf{with} generative augmentations}
    \end{subfigure}

    \caption{Evaluation of automatic EF on the \textbf{HUNT4 EF set} obtained via segmentation models trained with and without generative augmentations. The orange dots represent exams where at least one frame used in the calculation is outside the normal range for HUNT4 (depth $>150$mm or sector angle $>70^\circ$). The reference EF values are obtained using EchoPAC software (GE HealthCare).}
    \label{fig: ef}
\end{figure*}

\begin{figure*}[h]
    \centering
    \begin{subfigure}{0.382\linewidth}
        \centering
        \includegraphics[trim={1.5cm 0cm 1.5cm 0cm},clip,width=\textwidth]{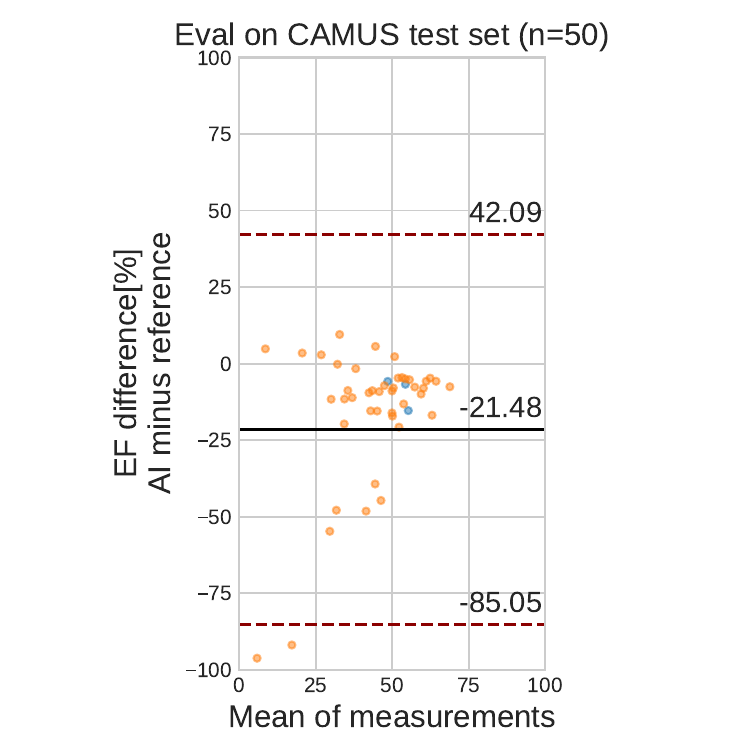}
        \caption{Trained on \textbf{HUNT4} \textbf{without} generative augmentations}
    \end{subfigure}
    \hspace{1.5cm}
    \begin{subfigure}{0.382\linewidth}
        \centering
        \includegraphics[trim={1.5cm 0cm 1.5cm 0cm},clip,width=\textwidth]{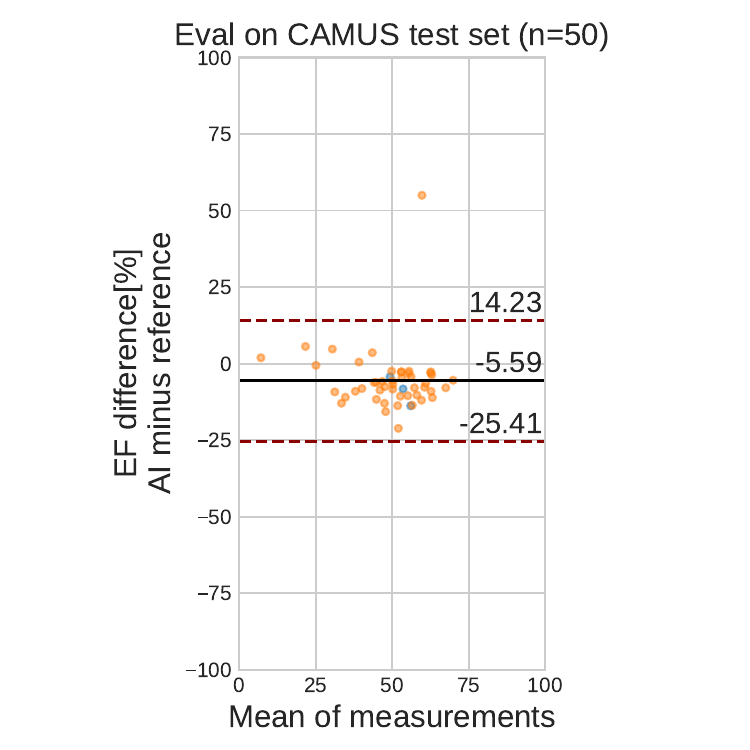}
        \caption{Trained on \textbf{HUNT4} \textbf{with} generative augmentations}
    \end{subfigure}


    \begin{subfigure}{0.382\linewidth}
        \centering
        \includegraphics[trim={1cm 1.6cm 0.4cm 2cm},clip,width=\textwidth]{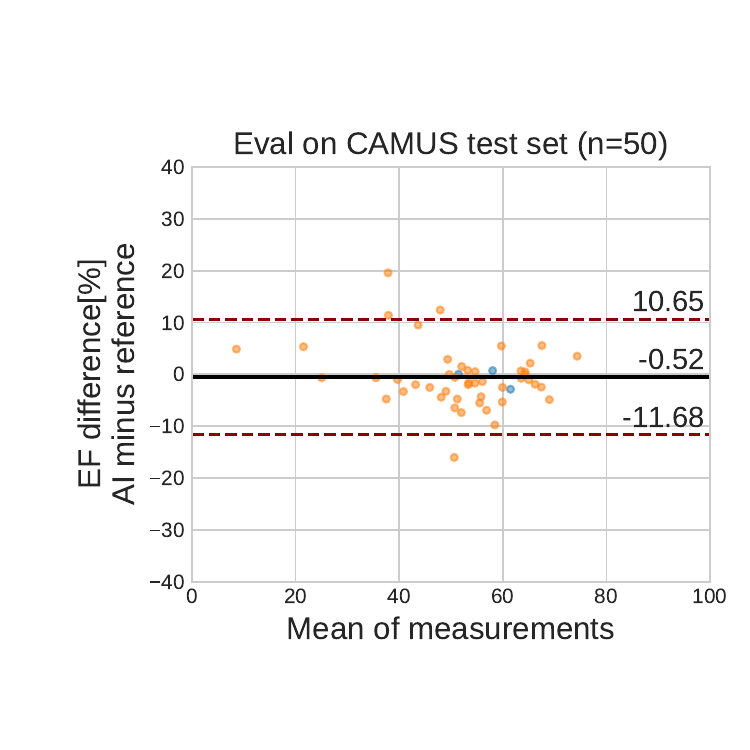}
        \caption{Trained on \textbf{CAMUS} \textbf{without} generative augmentations}
    \end{subfigure}
    \hspace{1.5cm}
    \begin{subfigure}{0.382\linewidth}
        \centering
        \includegraphics[trim={1cm 1.6cm 0.4cm 2cm},clip,width=\textwidth]{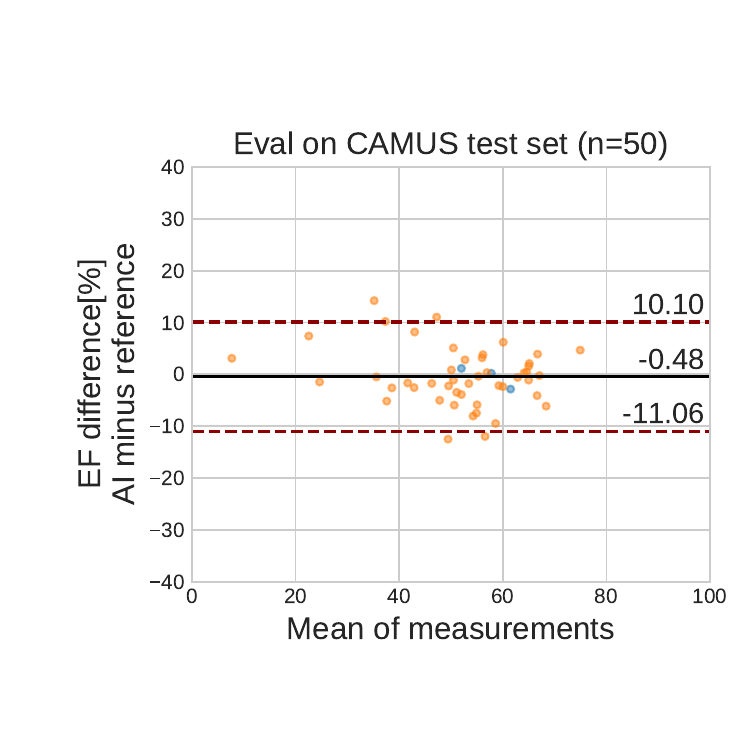}
        \caption{Trained on \textbf{CAMUS} \textbf{with} generative augmentations}
    \end{subfigure}

    \caption{Evaluation of automatic EF on \textbf{CAMUS} obtained via segmentation models trained with and without generative augmentations. The orange dots represent exams where at least one frame used in the calculation is outside the normal range for HUNT4 (depth $>150$mm or sector angle $>70^\circ$). The reference EF values are obtained using the automatic EF algorithm \cite{smistad2020real, van2024towards} with the reference segmentation masks.}
    \label{fig: ef}
\end{figure*}

\end{document}